\documentclass[aps,reprint,12pt]{revtex4}

\usepackage{commath}
\usepackage{amsmath}
\usepackage{graphicx}
\usepackage{hyperref}
\usepackage{txfonts}

\draft 

\begin{document}


\title{Optical excitation of surface plasma waves without grating structures}


\author{Hai-Yao Deng$^{1}$}
\email{haiyao.deng@gmail.com}
\author{Feng Liu$^{1}$}
\email{ruserzzz@gmail.com} 
\author{Katsunori Wakabayashi$^{1,2}$}
\email{waka@kwansei.ac.jp}
\affiliation{$^1$Department of Nanotechnology for Sustainable Energy,
School of Science and Technology, Kwansei Gakuin University, Gakuen 2-1,
Sanda 669-1337, Japan}
\affiliation{$^2$National Institute for Materials Science (NIMS), Namiki 1-1, Tsukuba 305-0044, Japan}


\begin{abstract} 
 Surface plasma waves (SPWs) are usually discussed in the context
 of a metal in contact with a dielectric. However, they can also exist
 between two metals. In this work we study these bimetallic waves. We find that their dispersion curve always cuts the light line, which allows direct optical coupling
 without surface grating structures. We propose practical schemes to excite them and the excitation efficiency is estimated. We also show that these waves can be much less lossy than conventional SPWs and their losses can be systematically controlled, a highly desirable attribute in applications. Conducting metal oxides are apt for experimental studies. 
\end{abstract}


\maketitle 

Surface plasma waves (SPWs) are widely known as charge density undulations propagating at the interface between a metal and a dielectric (such as vacuum)~\cite{ritchie1957plasma,economu,maier2007plasmonics,pitarke2007theory}. Due to highly localized electromagnetic (EM) fields associated with them, SPWs have seen applications in many areas such as Raman scattering~\cite{xu2013near,berini}, near-field spectroscopy~\cite{berndt1991inelastic} and bio-sensing~\cite{chien2006direct,hoa2007towards,rodrigo2015mid}. In the past decade or so, SPWs have emerged as a pivotal player in sub-wavelength optics~\cite{barnes2003surface,fakonas2015path}, where it is necessary to render efficient coupling between light (propagating EM waves) and SPWs. As the wavelength of SPW is much shorter than that of light at the same frequency (momentum mismatch), the desired coupling can be achieved only via contrived surface structures such as grating~\cite{maier2007plasmonics}. In this Letter, we investigate SPWs supported at the interface between two metals and show that such SPWs can couple to light without gratings. Although these waves were noticed in early studies~\cite{Stern1960,f,k,apell} and are garnering interest in the field of core-shell nano-particles~\cite{Zhu,chau2006}, a comprehensive understanding, especially of their coupling with light, has yet to emerge~\cite{raether1988}. We systematically look into their properties and propose practical while efficient means of launching them.

\begin{figure*}
\begin{center}
\includegraphics[width=0.95\textwidth]{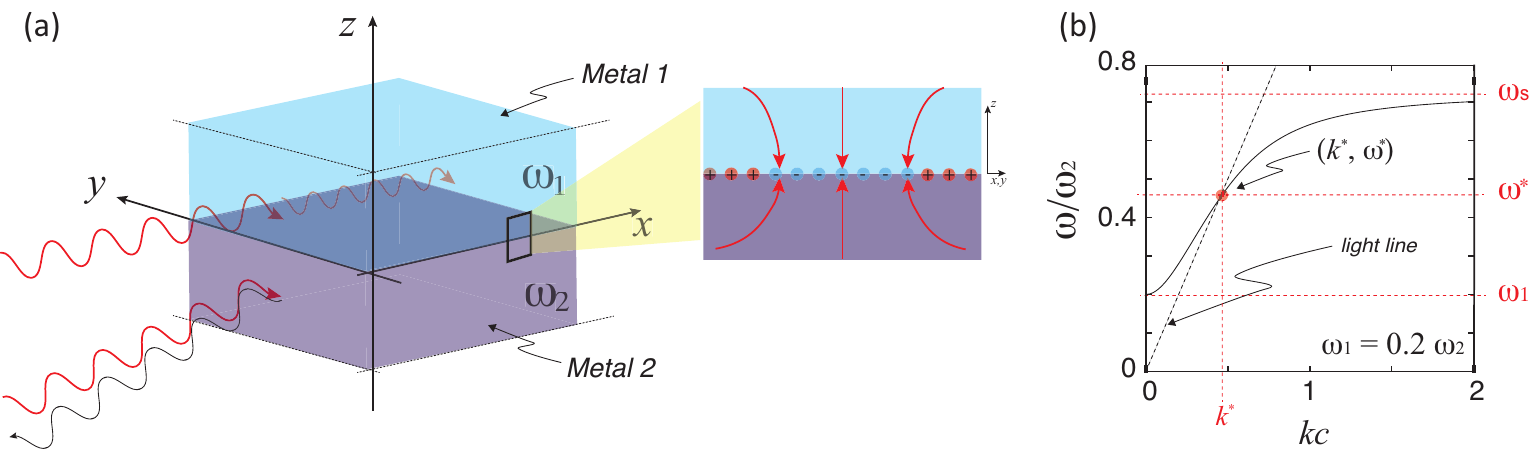}
\end{center}
\caption{SPWs supported at a bimetallic interface can be optically excited without the need of grating structures. (a) Setup of the system: two metals characterized by their bulk plasma frequencies $\omega_j$, where $j=1,2$ labels the metals, touch at a planar interface $z=0$. SPWs can propagate along e.g. $x$-direction with wavenumber $k$. When the electronic mean free paths in both metals are much smaller than $2\pi/k$, SPWs may be represented by charges completely localized at the interface. Arrowed thick lines indicate the electric field due to those charges. (b) Dispersion relation $\omega(k)$, assuming $\omega_2>\omega_1$. SPWs can exist only in the window $[\omega_1,\omega_s]$, where $\omega^2_s = (\omega^2_1+\omega^2_2)/2$. The light line inevitably cuts the SPW dispersion at $(k^*,\omega^*)$. Here $\omega^*=\sqrt{\omega_1\omega_2}$ for light traveling in vacuum but shifts to elsewhere for light propagating in other media, allowing SPWs to be excited in the entire spectrum directly with light, as illustrated in (a), where a beam of TM polarized light incident at the interface enters metal $1$ and hence generates SPWs. The maximal efficiency as estimated by Eq.~(\ref{7}) peaks where the crossing occurs, which is readily advocated by simulations (see Fig.~\ref{figure:f2}).\label{figure:f1}}
\end{figure*}

We consider a planar interface between two metals each characterized by their bulk plasma frequency $\omega_j$, as shown in Fig.~\ref{figure:f1} (a). Here $j=1,2$ labels the metals. For the sake of definiteness, we assume $\omega_2>\omega_1$ in this Letter. SPWs can be supported at the interface. Unlike in ordinary metal-vacuum structures, however, the dispersion relation $\omega(k)$ of these SPWs is not photon-like in the long wavelength limit. Instead, as their wavenumber $k$ increases from zero, their frequency $\omega$ rises from $\omega_1$, the lower of the bulk plasma frequencies of the metals, and approaches an upper limit. As such, the dispersion relation of light always intercepts $\omega(k)$, thereby allowing SPWs to be directly excited with light without grating, which makes the key observation of this Letter. 

At first thought, one might think that such waves must be extremely lossy and thus of limited practical uses. On the contrary, we find that bimetallic SPWs can be much less lossy and their losses can be controlled around the point of interception. 

We first describe the full dispersion relation $\omega(k)$ as plotted in Fig.~\ref{figure:f1} (b), where we see that SPWs are admitted only with frequency $\omega\in[\omega_1,\omega_s]$. The upper bound $\omega_s = \sqrt{(\omega^2_1+\omega^2_2)/2}$ was first reported in Ref.~\cite{Stern1960} and it reduces to the usually quoted SPW frequency~\cite{pitarke2007theory} if one of the metals is replaced by a dielectric, i.e. by setting $\omega_1=0$. This bound arises in the electrostatic limit, where the speed of light (in vacuum) $c$ can be taken as infinite. The lower bound $\omega_1$ is achieved at $k=0$. Now that the frequency of light runs over the entire real axis, the light line must traverse the SPW line at $(k^*,\omega^*)$. The value of $(k^*,\omega^*)$ depends on the speed of light. For light traveling in vacuum, it is given by 
\begin{equation}
\omega^* = \sqrt{\omega_1\omega_2}, ~ k^* = \omega^*/c,
\label{2}
\end{equation}
which is indicated in Fig.~\ref{figure:f1} (b). For light traveling in other media, the light line has a different slope and hence the crossing takes place elsewhere.

The reason why $\omega_1$ sets the lower bound for the SPW frequency is easy to understand.
Since SPWs harbor charges at the interface, the normal component of the electric field changes sign across the interface. To ensure the continuity of the normal component of the electric displacement field, the dielectric function must change its sign. The dielectric function of ideal metals is completely determined by its bulk plasma frequency if the SPW wavelength is the only relevant length scale in question as assumed here. For metal $j$ it is given by $\epsilon_j(\omega) = 1-\omega^2_j/\omega^2$. Now we must have $\epsilon_1(\omega)\cdot \epsilon_2(\omega)<0$, which yields $\omega>\omega_1$. This indicates that metal $1$ effects as a dielectric, since $\epsilon_1(\omega)>0$.   

The full profile of $\omega(k)$ exhibited in Fig.~\ref{figure:f1} (b) can be established with the knowledge that the interfacial charges feel an effective dielectric function given by 
\begin{equation}
\epsilon(\omega) = \epsilon_1(\omega)\cdot\epsilon_2(\omega)/\left(\epsilon_1(\omega)+\epsilon_2(\omega)\right).\label{0}
\end{equation}
This means the EM waves carried by SPWs must bear this dispersion relation: $k = (\omega/c)\sqrt{\epsilon(\omega)}$, which actually describes TM-polarized solutions of Maxwell's equations applied to our system~\cite{maier2007plasmonics}. Solving this equation reproduces Fig.~\ref{figure:f1} (b). SPW solutions exist only if $\epsilon(\omega)>0$ so that $k$ is real, thereby establishing $\omega_1<\omega<\omega_s$. The electrostatic limit is attained for $\epsilon_1+\epsilon_2=0$. The light line and the SPW line cross where $\epsilon(\omega^*)=1$, which retrieves Eq.~(\ref{2}). For light traveling in a media of refractive index $n$, the crossing occurs where $\epsilon(\omega^*)=n^2$. 

We can also derive the dispersion relation $\omega(k)$ by studying the motion of charges rather than the EM fields. The conductivity for our system can be written as $\sigma(z;\omega) = i\omega^2(z)/4\pi\omega$, where $\omega(z)$ takes the value of $\omega_1$ ($\omega_2$) for $z\geq0$ ($z<0$). The presence of a charge density $\rho(\vec{x})e^{-i\omega t}$ produces an electric field $\vec{E}(\vec{x})e^{-i\omega t}$, which then drives a current of density $\vec{J}(\vec{x})e^{-i\omega t}$ with $\vec{J}(\vec{x}) = \sigma(z;\omega)\vec{E}(\vec{x})$. Applying the equation of continuity to $\rho$ and $\vec{J}$ and using $\partial_z\omega^2(z)=\delta(z)(\omega^2_1-\omega^2_2)$, where $\delta(z)$ is the Dirac function, we get
\begin{equation}
 \left(\omega^2-\omega^2(z)\right)\rho(\vec{x}) = \delta(z)\left(E_z(0)/2\pi\right)(\omega^2_1-\omega^2_2)/2.
  \label{3}
\end{equation}
SPWs are represented by solutions with non-vanishing $E_z(0)$, for which the charge density is completely localized at the interface: $\rho(\vec{x})=\rho_s(x,y)\delta(z)$, where $\rho_s(x,y)$ denotes interfacial charge density. Taking $\rho_s(x,y)=\rho_se^{ikx}$, we find $E_z(0)=-2\pi\rho_s$ in the electrostatic limit. Inserting this in Eq.~(\ref{3}), we arrive
\begin{equation}
 1 = (1/2)(\omega^2_2-\omega^2_1)/(\omega^2-\omega^2_1).
  \label{4}
\end{equation}
This Equation can be extended beyond the electrostatic limit to reproduce the full SPW spectrum, if $E_z(0)$ in Eq.~(\ref{3}) is evaluated with the inclusion of retardation effects~\cite{deng2015retardation}, though the charge density is still completely localized at the interface~\footnote[2]{This is because the conductivities used in the calculation encode only local responses. See \cite{deng} for more account.}. This derivation makes it clear that SPWs originate from an abrupt change in conductivity~\cite{deng}. 

\begin{figure}
\begin{center}
\includegraphics[width=0.47\textwidth]{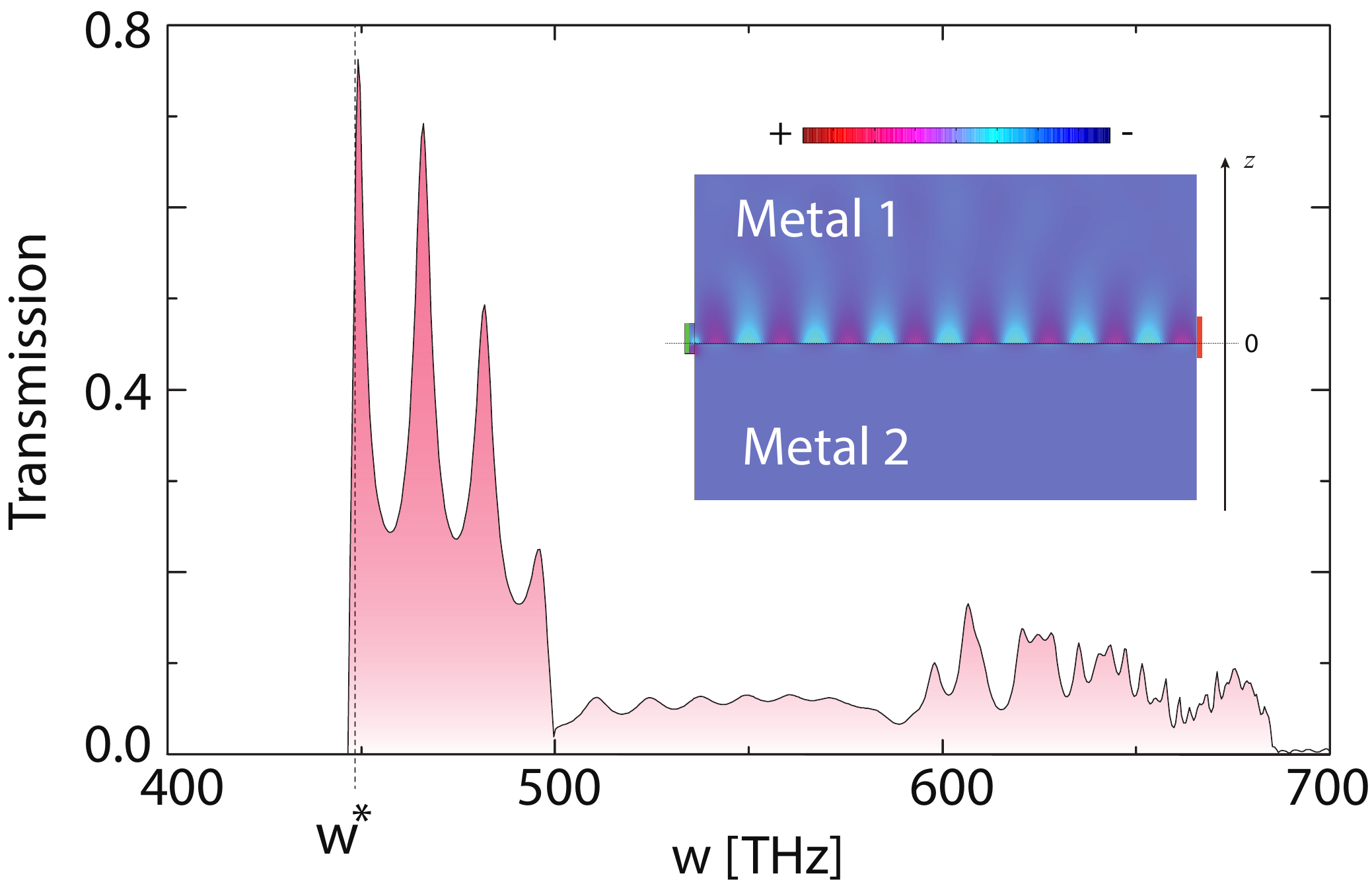}
\end{center}
\caption{Simulated SPW excitation efficiency. A wave is sent from a wave guide and impinges upon a bimetallic interface welded to the guide. SPWs are then excited. A simulation of this process is conducted using a commercial package COMSOL. The inset sketches the setup and exhibits a map of the electric field ($z$-component) of the excited SPW near $\omega^*$. The green (red) bar in the inset indicates the port for incoming (out-going) waves, and the transmission is calculated as the ratio between the energy fluxes received at the ports. \label{figure:f2}}
\end{figure}

The advantage of SPWs between metals lies in the possibility of being optically excited without grating structures, thanks to the unavoidable crossing of the light line and the SPW dispersion. In Fig.~\ref{figure:f1} (a) we conceive a scheme to implement this, in which a beam of TM polarized light at frequency $\omega^*$ shines sideways upon the interface from a media of refractive index $n$. Although it cannot enter metal $2$, this beam does get into metal $1$ because $\omega^*>\omega_1$. As such, it can resonantly induce SPW oscillations at the crossing frequency $\omega^*$ determined by $\epsilon(\omega^*)=n^2$. To estimate the excitation efficiency, let us write the incoming wave as $U(x,z,t) = u(z)e^{i(kx-\omega t)}$, where $U$ stands for a component of the EM fields. The general scattering problem by the structure is hard to solve. However, if we prepare the wave in such a way that $u(z)$ exactly matches the profile of that of the SPW ($u_{spw}$), the problem can be rigorously solved. In such cases, the incident incoming wave will be partially reflected back with reflection coefficient $R$ and partially transmitted in the form of a SPW with coefficient $T=1-R$. Matching these waves at the boundary gives 
\begin{equation}
R = |(\sqrt{\epsilon(\omega)}-n)/(\sqrt{\epsilon(\omega)}+n)|^2,\label{7}
\end{equation} 
where $\epsilon(\omega)$ is given by Eq.~(\ref{0}). This expression has the same form of Fresnel formula but bears obviously different contents. We then find $R=1$ and $T=0$ for $\omega=\omega_1, \omega_s$, while $R=0$ and $T=1$ for $\omega=\omega^*$. If $u(z)$ does not exactly fit the SPW profile, other modes than SPWs will be excited and this equation shall not apply. Then $T$ will be reduced. 

To further vindicate the idea, we have performed numerical simulations with the finite element method by a commercial package COMSOL. We set up the system consisting of two metals with bulk plasma frequencies $\omega_2=10^3$THz and $\omega_1=0.2\omega_2$, respectively. Then we send a wave (of frequency $\omega$) through a waveguide (of refractive index one) welded to the interface between the metals. We calculate the transmission as the ratio of the energy flux received at the red port to that at the green port (see the inset of Fig.~\ref{figure:f2}) as a function of $\omega\in[\omega_1,\omega_s]$. The result is displayed in Fig.~\ref{figure:f2}. We see that only in the neighborhood of $\omega^*$, where the light line cuts the SPW dispersion, the energy can be well transmitted and thus SPWs be excited, in good agreement with Eq.~(\ref{7}). Note that the transmission is remarkably high, achieving nearly $80\%$ around the resonance point. 

Since the variation of the refractive index $n$ of the incidence media can tune the slope of the light line [see Fig.~\ref{figure:f1} (b)], $\omega^*$ can run all the way from $\omega_1$ to $\omega_s$. Thus, the whole spectrum of SPWs can in principle be excited.

The scheme can also be used to convert light into SPW with a different wavelength. We let a light beam transmit through a slab of dielectric (of index $n$) and then hit upon a bimetallic interface (with which the slab is in contact). SPWs are thus excited, whose wavelength, however, is only $1/n$ times that of the original incident light. The slab thickness should be chosen to maximize the transmission. This idea may find applications in nano-optics.    

\begin{figure}
\begin{center}
\includegraphics[width=0.45\textwidth]{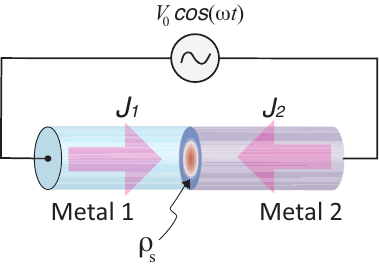}
\end{center}
\caption{Excitation of SPWs by AC circuits. As the system consists only of metals, the waves can be excited via an AC circuit driven by a voltage $V(t)=V_0\cos(\omega t)$. Assuming the interface is smaller than the SPW wavelength in dimension and treating it as a capacitor, the charge density $\rho_s$ can be shown with a resonance at $\omega_s$.The same resonance shows up also in the current densities $J_{1/2}$. Near $\omega_s$, $J_1$ and $J_2$ are opposite in direction. See Eqs.~(\ref{5}) and (\ref{6}). \label{figure:f3}}
\end{figure}

Since our system is merely composed of metals, we may also excite the waves using a simple AC circuit as sketched in Fig.~\ref{figure:f3}. The samples may be fabricated with sufficiently small cross-sections (less than a SPW wavelength) so that variations of charge density and currents can be neglected within the interface. The circuit can be solved by treating the interface as a capacitor in the electrostatic limit. The interfacial charge density varies as $\rho_s\cos(\omega t)$, where
\begin{equation}
 \rho_s = \left[(\omega^2_2-\omega^2_1)/(\omega^2-\omega^2_s)\right]\left(V_0/4\pi D\right).\label{5}
\end{equation}
Here $D$ denotes the total length of the sample and $V_0$ the amplitude of the applied AC voltage. We see that $\rho_s$ develops a resonance at the SPW frequency. The current density is given by
\begin{equation}
J_{1/2} = \left(\omega^2_{1/2}\frac{\sin\omega t}{\omega}\right)\left[1\pm(1/2)(\omega^2_2-\omega^2_1)/(\omega^2_s-\omega^2)\right]\frac{V_0}{D},\label{6}
\end{equation}
which shows the same resonance. Near the resonance, the current flows in opposite directions across the interface, as pictured in Fig.~\ref{figure:f3}. Note that bulk waves cannot be excited this way, as is evident from Eqs.~(\ref{5}) and (\ref{6}).

\begin{figure}
\begin{center}
\includegraphics[width=0.47\textwidth]{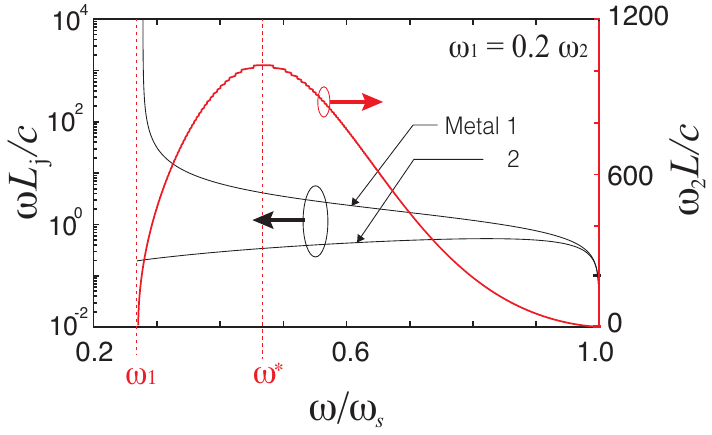}
\end{center}
\caption{Various length scales of practical importance. $L_j$ is the penetration depth of the electric fields in the $j$-th metal. Clearly, $L_2<c/\omega$ while $L_1\gg L_2$. The profile of $L_1$ ($L_2$) is analogous to that on the dielectric (metal) side of conventional SPWs. $L$ is the propagation distance, which peaks at $\omega^*$. In computing $L$, we have chosen $\gamma_1=\gamma_2=0.01\omega$, where $\gamma_j$ is the electronic relaxation rate.\label{figure:f4}}
\end{figure}

In practical uses of SPWs, apart from the wavelength $2\pi/k$, there are three other important lengths, which we denote by $L_1$, $L_2$ and $L$, respectively~\cite{1464-4258-8-4-S06}. They are all plotted in Fig.~\ref{figure:f4}. The penetration depth, $L_j=1/\sqrt{k^2-(\omega/c)^2\epsilon_j}$, describes how far the associated EM fields extend into the bulk of the $j$-th metal. As we can see in the simulations (Fig.~\ref{figure:f2}), the EM fields are much more strongly localized on the side of metal $2$, in line with the fact that $L_2\ll L_1$, as shown in Fig.~\ref{figure:f4}. In fact, $L_2$ is less than the light wavelength $2\pi c/\omega$ at all $\omega$ while $L_1$ is almost always larger. The trend of $L_1(\omega)$ [$L_2(\omega)$] resembles that on the dielectric (metal) side in conventional SPWs, again suggesting that metal $1$ acts as a dielectric. 

The length $L$ is critical in using SPWs to carry photons. It describes how far a SPW wave packet can propagate before it damps off due to Joule heat. We may write $L=v_g\tau$, where $\tau$ is the SPW lifetime and $v_g = d\omega/dk = c/ \left(\sqrt{\epsilon(\omega)} + d\sqrt{\epsilon(\omega)}/{d\ln\omega}\right)$ denotes its group velocity. 

The SPW lifetime $\tau$ quantifies the losses experienced by the waves. At first thought one might think that bimetallic SPWs must be heavily lossy due to the presence of an extra metal, in comparison with dielectric/metal SPWs. Contrary to this expectation, we show that they can actually be much less lossy. For this purpose, we evaluate $\tau$ within the Drude model. The validity of this model and the calculation details are discussed in Ref.~\footnote[1]{Supplemental Online Information.}. Here we only quote the result of $\tau$ for the crossing point $(k^*,\omega^*)$, which is of special interest to us. It is given by
\begin{equation}
\tau^{-1} \approx \frac{\gamma_1+\gamma_2}{2}~\frac{\omega_1}{\omega_2}~\frac{1}{(1-\omega_1/\omega_2)^2}, 
\end{equation}
where $\gamma_j$ denotes the electronic relaxation rate in metal $j$. This expression shows that $\tau$ can be systematically improved by choosing suitable materials so that $\omega_1/\omega_2$ is small. However, for conventional SPWs, $\tau^{-1}$ can be considerably larger. For example, for a vacuum/metal structure, we find~\cite{Note1} $\tau^{-1} \approx (\gamma/2)(\omega^2/\omega^2_p)/[(1-\omega^2/\omega^2_p)(1-2\omega^2/\omega^2_p)]$, which would diverge in the non-retarded regime where $\omega\sim\omega_p/\sqrt{2}$. Here $\gamma$ and $\omega_p$ are the relaxation rate and characteristic frequency of the metal, respectively. 

Consequently the propagation distance $L$ can be considerably larger than light wavelength $2\pi c/\omega$. In Fig.~\ref{figure:f4} we plot $L$ against frequency for $\gamma_1=\gamma_2=0.01\omega$, which is typical for most plasmonic materials~\cite{west,johnson1972}. Note that $L(\omega)$ peaks at $\omega^*$ with a value one thousand times as large as the light wavelength, an attribute highly wanted in applications. 

In a recent work Bliokh \textit{et al.} showed that conventional vacuum-metal SPWs display an optical analog of quantum spin Hall effect~\cite{Bliokh26062015}, where the spin $\bar{S}$ carried by SPWs is locked to the wave vector $k$, i.e., reversing $k$ results in reversed $\bar{S}$. Obviously, this property is also possessed by bimetallic SPWs, because these waves are described by the same equations, i.e. Eqs. (4) - (6) in Ref.~\cite{Bliokh26062015}, as the conventional SPWs. Moreover, these authors ascribed the existence of conventional SPWs to a topological difference between the optical media involved. The optical spin Chern number $C_{\sigma}$ for vacuum ($C_{\sigma}\neq 0$) differs from that for optically opaque metals ($C_{\sigma}=0$). This connection can as well be established for bimetallic SPWs, because transparent metals are topologically equivalent to the vacuum. Actually, in order to have nonzero $C_{\sigma}$, one only needs the existence of the $k$-sphere (see Ref.~\cite{Bliokh26062015} for definition). For light in metals we have $k^2 = (\omega^2-\omega^2_p)/c^2$, where $\omega_p$ denotes the bulk plasma frequency. Thus, the $k$-sphere exists if and only if $\omega>\omega_p$, i.e. when the metal is transparent. Returning to bimetallic SPWs, we see that metal $1$ is transparent while metal $2$ is opaque for $\omega\in(\omega_1,\omega_s)$, suggesting the same geometrical origin of these waves.

Being less lossy and directly excitable, bimetallic SPWs are evidently favorable with plasmonics. In experimental studies the challenge is to find suitable materials. Ideally, the conductors in the bimetallic structure need be well separated in their bulk plasma frequencies, i.e. $\omega_1/\omega_2\ll 1$. In the meanwhile, inter-band transitions need be avoided in the region of frequencies of interest, i.e., $\omega^*$ should stay far off the inter-band transition threshold of either constituent material. Noble metals like gold and silver, though most commonly experimented with, may not be good candidates, because their bulk frequencies lie very close ($\sim 9$ eV) and inter-band transitions might prevail at frequencies thereof~\cite{west,drachev2008}. On the other hand, conducting metal oxides such as indium tin oxides (ITO) and zinc oxides (e.g. Al:ZnO) can be very apt for studying bimetallic SPWs~\cite{west,rhodes2006,losego2009}. These materials have attracted plenty of attention in recent years for their widely tunable plasma frequencies (from infrared to ultra-violet), low losses (e.g. relaxation rate could be lower than that of silver) and clearance of inter-band transitions~\cite{losego2009}. Therefore, SPWs in bimetallic structures composed of these materials should be well described by our theory. 

We have thus studied SPWs at bimetallic interfaces. These waves have a dispersion inevitably traversing the light line and they can be resonantly excited without grating structures. Also, they can be much less lossy near the crossing points. These features can not exist in dielectric/metal or cladding structures~\cite{economu,prada} and they are highly desired in current plasmonics. We hope the present work will stimulate further theoretical and experimental studies on these waves. 
 
\textit{Acknowledgment} -- HYD acknowledges the International Research
Fellowship of the Japan Society for the Promotion of Science
(JSPS). This work is supported by JSPS KAKENHI Grant No. 15K13507 and
MEXT KAKENHI Grant No. 25107005.

\newpage

\appendix
\section{Losses of bimetallic surface plasma waves}

Losses constitute a very important issue in making use of SPWs. One might naively suppose that SPWs between bimetallic structures must be heavily lossy, in comparison with e.g. SPWs in dielectric/metal structures, because introducing one more metal is likely to cause additional losses. However, here we show that this is not so. We find that the losses with bimetallic SPWs can be systematically reduced. This finding is in accord with the large propagation distance shown in Fig. 4 in the main manuscript. 

Our discussions will be framed within the Drude model, by which we approximate the dielectric function of a metal as 
\begin{equation}
\epsilon_j(\omega) = 1-\frac{\omega^2_j}{\omega(\omega+i\gamma_j)}=1-\frac{\omega^2_j}{\omega^2}\frac{1}{1+i\delta_j} = \epsilon'_j(\omega)+i\epsilon''(\omega), \quad \gamma_j = \delta_j \omega.\label{a1}
\end{equation}
Here $j$ labels the metal and $\gamma_j>0$ denotes the electronic relaxation rate. One should note that the validity of this model holds even when losses arise due to inter-band transitions, as long as such transitions do not modify too much the real part of $\epsilon_j(\omega)$, in which case inter-band transitions can be taken care of by the relaxation rate $\gamma_j$. Therefore, it is expected to be valid for frequencies not so close to the inter-band transition threshold. For noble metals, it is very accurate for $\omega$ below $\sim 2.6$ eV, while for conducting metal oxides, it is applicable at even higher frequencies. 

The equation determining the spectrum of SPWs is the following:
\begin{equation}
\frac{k^2c^2}{\omega^2} = \epsilon(\omega),\quad \epsilon(\omega) = \frac{\epsilon_1(\omega)\epsilon_2(\omega)}{\epsilon_1(\omega)+\epsilon_2(\omega)}. \label{a2}
\end{equation}
Due to finite $\gamma_j$, $\epsilon(\omega)$ becomes complex and $\omega$ must also be complex. So we put $\omega = \Omega - i\Gamma$, where both $\Omega$ and $\Gamma$ are real. We shall assume $\delta = \Gamma/\Omega$ is small, which is reasonable if $\delta_j$ is small. Then we can rewrite Eq.~(\ref{a2}) as follows,
\begin{equation}
\frac{(\Omega-i\Gamma)^2}{k^2c^2} \approx \frac{1}{\epsilon(\Omega)}, \label{a3}
\end{equation} 
from which we find 
\begin{equation}
\frac{\Omega^2}{k^2c^2} \approx \frac{1}{1-\omega^2_1/\Omega^2}+\frac{1}{1-\omega^2_2/\Omega^2} \label{a4}
\end{equation}
yielding $\Omega(k)\in[\omega_1,\omega_s]$ as plotted in Fig. 1 (c), and
\begin{equation}
\frac{2\Omega\Gamma}{k^2c^2} \approx \Im\left({\frac{1}{\epsilon(\Omega)}}\right)\label{a5}
\end{equation}
producing
\begin{equation}
\Gamma = \frac{1}{2}\left(\frac{\gamma_1
\omega^2_1/\Omega^2}{(1-\omega^2_1/\Omega^2)^2}+\frac{\gamma_2\omega^2_2/\Omega^2}{(1-\omega^2_2/\Omega^2)^2}\right)\frac{k^2c^2}{\Omega^2}.\label{a6}
\end{equation}
Of special interest to us concerns $\Gamma$ at $\omega^* = \sqrt{\omega_1\omega_2}$, where the light line cuts the SPW dispersion curve. There we get
\begin{equation}
\Gamma^* = \Gamma(\omega^*) = \frac{\gamma_1+\gamma_2}{2} ~\frac{\omega_1\omega_2}{(\omega_2-\omega_1)^2} = \frac{\gamma_1+\gamma_2}{2}~\frac{\omega_1}{\omega_2}~\frac{1}{(1-\omega_1/\omega_2)^2}. \label{a7}
\end{equation}
This expression shows that $\Gamma^*$ can be greatly reduced by choosing suitable materials so that $\omega_1/\omega_2$ is sufficiently small. In other words, the losses can be systematically controlled. 

To facilitate the comparison, let us also derive the losses for vacuum/metal SPWs. Here only one metal is involved. Let $\omega_p$ be its bulk plasma frequency and $\gamma$ the relaxation rate. In a similar fashion, we obtain for such waves $\omega = \Omega-i\Gamma$ with 
\begin{equation}
(\Omega/kc)^2 = \frac{2\Omega^2-\omega^2_p}{\Omega^2-\omega^2_p}
\end{equation}
and 
\begin{equation}
\Gamma = \frac{\gamma}{2} ~\frac{\Omega^2/\omega^2_p}{(1-\Omega^2/\omega^2_p)(1-2\Omega^2/\omega^2_p)}. 
\end{equation}
Here $\Gamma$ increases with $\Omega$ very quickly and the waves become extremely lossy (i.e., $\Gamma$ diverges) when $\Omega$ approaches $\omega_p/\sqrt{2}$. There is no way to control $\Gamma$ except to make better materials. 

Our analysis shows that bimetallic SPWs can actually be much less lossy than conventional SPWs, in contrast to the naive expectation. Therefore, such waves offer great advantages in plasmonics.

\bibliographystyle{apsrev4-1}
\bibliography{dengbib}

\begin{thebibliography}{31}%
\makeatletter
\providecommand \@ifxundefined [1]{%
 \@ifx{#1\undefined}
}%
\providecommand \@ifnum [1]{%
 \ifnum #1\expandafter \@firstoftwo
 \else \expandafter \@secondoftwo
 \fi
}%
\providecommand \@ifx [1]{%
 \ifx #1\expandafter \@firstoftwo
 \else \expandafter \@secondoftwo
 \fi
}%
\providecommand \natexlab [1]{#1}%
\providecommand \enquote  [1]{``#1''}%
\providecommand \bibnamefont  [1]{#1}%
\providecommand \bibfnamefont [1]{#1}%
\providecommand \citenamefont [1]{#1}%
\providecommand \href@noop [0]{\@secondoftwo}%
\providecommand \href [0]{\begingroup \@sanitize@url \@href}%
\providecommand \@href[1]{\@@startlink{#1}\@@href}%
\providecommand \@@href[1]{\endgroup#1\@@endlink}%
\providecommand \@sanitize@url [0]{\catcode `\\12\catcode `\$12\catcode
  `\&12\catcode `\#12\catcode `\^12\catcode `\_12\catcode `\%12\relax}%
\providecommand \@@startlink[1]{}%
\providecommand \@@endlink[0]{}%
\providecommand \url  [0]{\begingroup\@sanitize@url \@url }%
\providecommand \@url [1]{\endgroup\@href {#1}{\urlprefix }}%
\providecommand \urlprefix  [0]{URL }%
\providecommand \Eprint [0]{\href }%
\providecommand \doibase [0]{http://dx.doi.org/}%
\providecommand \selectlanguage [0]{\@gobble}%
\providecommand \bibinfo  [0]{\@secondoftwo}%
\providecommand \bibfield  [0]{\@secondoftwo}%
\providecommand \translation [1]{[#1]}%
\providecommand \BibitemOpen [0]{}%
\providecommand \bibitemStop [0]{}%
\providecommand \bibitemNoStop [0]{.\EOS\space}%
\providecommand \EOS [0]{\spacefactor3000\relax}%
\providecommand \BibitemShut  [1]{\csname bibitem#1\endcsname}%
\let\auto@bib@innerbib\@empty
\bibitem [{\citenamefont {Ritchie}(1957)}]{ritchie1957plasma}%
  \BibitemOpen
  \bibfield  {author} {\bibinfo {author} {\bibfnamefont {R.}~\bibnamefont
  {Ritchie}},\ }\href@noop {} {\bibfield  {journal} {\bibinfo  {journal} {Phys.
  Rev.}\ }\textbf {\bibinfo {volume} {106}},\ \bibinfo {pages} {874} (\bibinfo
  {year} {1957})}\BibitemShut {NoStop}%
\bibitem [{\citenamefont {Economou}(1969)}]{economu}%
  \BibitemOpen
  \bibfield  {author} {\bibinfo {author} {\bibfnamefont {E.~N.}\ \bibnamefont
  {Economou}},\ }\href@noop {} {\bibfield  {journal} {\bibinfo  {journal}
  {Phys. Rev.}\ }\textbf {\bibinfo {volume} {182}},\ \bibinfo {pages} {539}
  (\bibinfo {year} {1969})}\BibitemShut {NoStop}%
\bibitem [{\citenamefont {Maier}(2007)}]{maier2007plasmonics}%
  \BibitemOpen
  \bibfield  {author} {\bibinfo {author} {\bibfnamefont {S.~A.}\ \bibnamefont
  {Maier}},\ }\href@noop {} {\emph {\bibinfo {title} {Plasmonics: fundamentals
  and applications}}}\ (\bibinfo  {publisher} {Springer Science \& Business
  Media},\ \bibinfo {year} {2007})\BibitemShut {NoStop}%
\bibitem [{\citenamefont {Pitarke}\ \emph {et~al.}(2007)\citenamefont
  {Pitarke}, \citenamefont {Silkin}, \citenamefont {Chulkov},\ and\
  \citenamefont {Echenique}}]{pitarke2007theory}%
  \BibitemOpen
  \bibfield  {author} {\bibinfo {author} {\bibfnamefont {J.}~\bibnamefont
  {Pitarke}}, \bibinfo {author} {\bibfnamefont {V.}~\bibnamefont {Silkin}},
  \bibinfo {author} {\bibfnamefont {E.}~\bibnamefont {Chulkov}}, \ and\
  \bibinfo {author} {\bibfnamefont {P.}~\bibnamefont {Echenique}},\ }\href@noop
  {} {\bibfield  {journal} {\bibinfo  {journal} {Rep. Prog. Phys.}\ }\textbf
  {\bibinfo {volume} {70}},\ \bibinfo {pages} {1} (\bibinfo {year}
  {2007})}\BibitemShut {NoStop}%
\bibitem [{\citenamefont {Xu}\ \emph {et~al.}(2013)\citenamefont {Xu},
  \citenamefont {Li}, \citenamefont {Hasan}, \citenamefont {Ruoff},
  \citenamefont {Wang},\ and\ \citenamefont {Fan}}]{xu2013near}%
  \BibitemOpen
  \bibfield  {author} {\bibinfo {author} {\bibfnamefont {X.}~\bibnamefont
  {Xu}}, \bibinfo {author} {\bibfnamefont {H.}~\bibnamefont {Li}}, \bibinfo
  {author} {\bibfnamefont {D.}~\bibnamefont {Hasan}}, \bibinfo {author}
  {\bibfnamefont {R.~S.}\ \bibnamefont {Ruoff}}, \bibinfo {author}
  {\bibfnamefont {A.~X.}\ \bibnamefont {Wang}}, \ and\ \bibinfo {author}
  {\bibfnamefont {D.}~\bibnamefont {Fan}},\ }\href@noop {} {\bibfield
  {journal} {\bibinfo  {journal} {Adv. Fun. Mater.}\ }\textbf {\bibinfo
  {volume} {23}},\ \bibinfo {pages} {4332} (\bibinfo {year}
  {2013})}\BibitemShut {NoStop}%
\bibitem [{\citenamefont {Berini}\ and\ \citenamefont {Leon}(2012)}]{berini}%
  \BibitemOpen
  \bibfield  {author} {\bibinfo {author} {\bibfnamefont {P.}~\bibnamefont
  {Berini}}\ and\ \bibinfo {author} {\bibfnamefont {I.~D.}\ \bibnamefont
  {Leon}},\ }\href@noop {} {\bibfield  {journal} {\bibinfo  {journal} {Nat.
  Photonics}\ }\textbf {\bibinfo {volume} {6}},\ \bibinfo {pages} {16}
  (\bibinfo {year} {2012})}\BibitemShut {NoStop}%
\bibitem [{\citenamefont {Berndt}\ \emph {et~al.}(1991)\citenamefont {Berndt},
  \citenamefont {Gimzewski},\ and\ \citenamefont
  {Johansson}}]{berndt1991inelastic}%
  \BibitemOpen
  \bibfield  {author} {\bibinfo {author} {\bibfnamefont {R.}~\bibnamefont
  {Berndt}}, \bibinfo {author} {\bibfnamefont {J.~K.}\ \bibnamefont
  {Gimzewski}}, \ and\ \bibinfo {author} {\bibfnamefont {P.}~\bibnamefont
  {Johansson}},\ }\href@noop {} {\bibfield  {journal} {\bibinfo  {journal}
  {Phys. Rev. Lett.}\ }\textbf {\bibinfo {volume} {67}},\ \bibinfo {pages}
  {3796} (\bibinfo {year} {1991})}\BibitemShut {NoStop}%
\bibitem [{\citenamefont {Chien}\ and\ \citenamefont
  {Chen}(2006)}]{chien2006direct}%
  \BibitemOpen
  \bibfield  {author} {\bibinfo {author} {\bibfnamefont {F.-C.}\ \bibnamefont
  {Chien}}\ and\ \bibinfo {author} {\bibfnamefont {S.-J.}\ \bibnamefont
  {Chen}},\ }\href@noop {} {\bibfield  {journal} {\bibinfo  {journal} {Opt.
  Letters}\ }\textbf {\bibinfo {volume} {31}},\ \bibinfo {pages} {187}
  (\bibinfo {year} {2006})}\BibitemShut {NoStop}%
\bibitem [{\citenamefont {Hoa}\ \emph {et~al.}(2007)\citenamefont {Hoa},
  \citenamefont {Kirk},\ and\ \citenamefont {Tabrizian}}]{hoa2007towards}%
  \BibitemOpen
  \bibfield  {author} {\bibinfo {author} {\bibfnamefont {X.}~\bibnamefont
  {Hoa}}, \bibinfo {author} {\bibfnamefont {A.}~\bibnamefont {Kirk}}, \ and\
  \bibinfo {author} {\bibfnamefont {M.}~\bibnamefont {Tabrizian}},\ }\href@noop
  {} {\bibfield  {journal} {\bibinfo  {journal} {Biosensors and
  Bioelectronics}\ }\textbf {\bibinfo {volume} {23}},\ \bibinfo {pages} {151}
  (\bibinfo {year} {2007})}\BibitemShut {NoStop}%
\bibitem [{\citenamefont {Rodrigo}\ \emph {et~al.}(2015)\citenamefont
  {Rodrigo}, \citenamefont {Limaj}, \citenamefont {Janner}, \citenamefont
  {Etezadi}, \citenamefont {de~Abajo}, \citenamefont {Pruneri},\ and\
  \citenamefont {Altug}}]{rodrigo2015mid}%
  \BibitemOpen
  \bibfield  {author} {\bibinfo {author} {\bibfnamefont {D.}~\bibnamefont
  {Rodrigo}}, \bibinfo {author} {\bibfnamefont {O.}~\bibnamefont {Limaj}},
  \bibinfo {author} {\bibfnamefont {D.}~\bibnamefont {Janner}}, \bibinfo
  {author} {\bibfnamefont {D.}~\bibnamefont {Etezadi}}, \bibinfo {author}
  {\bibfnamefont {F.~J.~G.}\ \bibnamefont {de~Abajo}}, \bibinfo {author}
  {\bibfnamefont {V.}~\bibnamefont {Pruneri}}, \ and\ \bibinfo {author}
  {\bibfnamefont {H.}~\bibnamefont {Altug}},\ }\href@noop {} {\bibfield
  {journal} {\bibinfo  {journal} {Science}\ }\textbf {\bibinfo {volume}
  {349}},\ \bibinfo {pages} {165} (\bibinfo {year} {2015})}\BibitemShut
  {NoStop}%
\bibitem [{\citenamefont {Barnes}\ \emph {et~al.}(2003)\citenamefont {Barnes},
  \citenamefont {Dereux},\ and\ \citenamefont {Ebbesen}}]{barnes2003surface}%
  \BibitemOpen
  \bibfield  {author} {\bibinfo {author} {\bibfnamefont {W.~L.}\ \bibnamefont
  {Barnes}}, \bibinfo {author} {\bibfnamefont {A.}~\bibnamefont {Dereux}}, \
  and\ \bibinfo {author} {\bibfnamefont {T.~W.}\ \bibnamefont {Ebbesen}},\
  }\href@noop {} {\bibfield  {journal} {\bibinfo  {journal} {Nature}\ }\textbf
  {\bibinfo {volume} {424}},\ \bibinfo {pages} {824} (\bibinfo {year}
  {2003})}\BibitemShut {NoStop}%
\bibitem [{\citenamefont {Fakonas}\ \emph {et~al.}(2015)\citenamefont
  {Fakonas}, \citenamefont {Mitskovets},\ and\ \citenamefont
  {Atwater}}]{fakonas2015path}%
  \BibitemOpen
  \bibfield  {author} {\bibinfo {author} {\bibfnamefont {J.~S.}\ \bibnamefont
  {Fakonas}}, \bibinfo {author} {\bibfnamefont {A.}~\bibnamefont {Mitskovets}},
  \ and\ \bibinfo {author} {\bibfnamefont {H.~A.}\ \bibnamefont {Atwater}},\
  }\href@noop {} {\bibfield  {journal} {\bibinfo  {journal} {New J. Phys.}\
  }\textbf {\bibinfo {volume} {17}},\ \bibinfo {pages} {023002} (\bibinfo
  {year} {2015})}\BibitemShut {NoStop}%
\bibitem [{\citenamefont {Stern}\ and\ \citenamefont
  {Ferrell}(1960)}]{Stern1960}%
  \BibitemOpen
  \bibfield  {author} {\bibinfo {author} {\bibfnamefont {E.~A.}\ \bibnamefont
  {Stern}}\ and\ \bibinfo {author} {\bibfnamefont {R.~A.}\ \bibnamefont
  {Ferrell}},\ }\href@noop {} {\bibfield  {journal} {\bibinfo  {journal} {Phys.
  Rev.}\ }\textbf {\bibinfo {volume} {120}},\ \bibinfo {pages} {130} (\bibinfo
  {year} {1960})}\BibitemShut {NoStop}%
\bibitem [{\citenamefont {Apell}\ and\ \citenamefont
  {Lundqvist}(1984)}]{apell}%
  \BibitemOpen
  \bibfield  {author} {\bibinfo {author} {\bibfnamefont {P.}~\bibnamefont
  {Apell}}\ and\ \bibinfo {author} {\bibfnamefont {S.}~\bibnamefont
  {Lundqvist}},\ }\href@noop {} {\bibfield  {journal} {\bibinfo  {journal}
  {Physica Scripta}\ }\textbf {\bibinfo {volume} {30}},\ \bibinfo {pages} {360}
  (\bibinfo {year} {1984})}\BibitemShut {NoStop}%
\bibitem [{\citenamefont {Forstmann}\ and\ \citenamefont
  {Stenschke}(1978)}]{f}%
  \BibitemOpen
  \bibfield  {author} {\bibinfo {author} {\bibfnamefont {F.}~\bibnamefont
  {Forstmann}}\ and\ \bibinfo {author} {\bibfnamefont {H.}~\bibnamefont
  {Stenschke}},\ }\href@noop {} {\bibfield  {journal} {\bibinfo  {journal}
  {Phys. Rev. B}\ }\textbf {\bibinfo {volume} {17}},\ \bibinfo {pages} {1489}
  (\bibinfo {year} {1978})}\BibitemShut {NoStop}%
\bibitem [{\citenamefont {Henneberger}(1998)}]{k}%
  \BibitemOpen
  \bibfield  {author} {\bibinfo {author} {\bibfnamefont {K.}~\bibnamefont
  {Henneberger}},\ }\href@noop {} {\bibfield  {journal} {\bibinfo  {journal}
  {Phys. Rev. Lett.}\ }\textbf {\bibinfo {volume} {80}},\ \bibinfo {pages}
  {2889} (\bibinfo {year} {1998})}\BibitemShut {NoStop}%
\bibitem [{\citenamefont {Zhu}(2009)}]{Zhu}%
  \BibitemOpen
  \bibfield  {author} {\bibinfo {author} {\bibfnamefont {J.}~\bibnamefont
  {Zhu}},\ }\href@noop {} {\bibfield  {journal} {\bibinfo  {journal} {Nanoscale
  Res. Lett.}\ }\textbf {\bibinfo {volume} {4}},\ \bibinfo {pages} {977}
  (\bibinfo {year} {2009})}\BibitemShut {NoStop}%
\bibitem [{\citenamefont {Chau}\ and\ \citenamefont
  {Elezzabi}(2006)}]{chau2006}%
  \BibitemOpen
  \bibfield  {author} {\bibinfo {author} {\bibfnamefont {K.~J.}\ \bibnamefont
  {Chau}}\ and\ \bibinfo {author} {\bibfnamefont {A.~Y.}\ \bibnamefont
  {Elezzabi}},\ }\href@noop {} {\bibfield  {journal} {\bibinfo  {journal}
  {Phys. Rev. B}\ }\textbf {\bibinfo {volume} {73}},\ \bibinfo {pages} {085419}
  (\bibinfo {year} {2006})}\BibitemShut {NoStop}%
\bibitem [{\citenamefont {Raether}(1988)}]{raether1988}%
  \BibitemOpen
  \bibfield  {author} {\bibinfo {author} {\bibfnamefont {H.}~\bibnamefont
  {Raether}},\ }\href@noop {} {\emph {\bibinfo {title} {Surface plasmons on
  smooth surfaces}}}\ (\bibinfo  {publisher} {Springer},\ \bibinfo {year}
  {1988})\BibitemShut {NoStop}%
\bibitem [{\citenamefont {Deng}\ and\ \citenamefont
  {Wakabayashi}(2015)}]{deng2015retardation}%
  \BibitemOpen
  \bibfield  {author} {\bibinfo {author} {\bibfnamefont {H.-Y.}\ \bibnamefont
  {Deng}}\ and\ \bibinfo {author} {\bibfnamefont {K.}~\bibnamefont
  {Wakabayashi}},\ }\href@noop {} {\bibfield  {journal} {\bibinfo  {journal}
  {Phys. Rev. B}\ }\textbf {\bibinfo {volume} {92}},\ \bibinfo {pages} {045434}
  (\bibinfo {year} {2015})}\BibitemShut {NoStop}%
\bibitem [{Note2()}]{Note2}%
  \BibitemOpen
  \bibinfo {note} {This is because the conductivities used in the calculation
  encode only local responses. See \cite {deng} for more account.}\BibitemShut
  {Stop}%
\bibitem [{\citenamefont {Deng}\ \emph {et~al.}(2015)\citenamefont {Deng},
  \citenamefont {Wakabayashi},\ and\ \citenamefont {Lam}}]{deng}%
  \BibitemOpen
  \bibfield  {author} {\bibinfo {author} {\bibfnamefont {H.-Y.}\ \bibnamefont
  {Deng}}, \bibinfo {author} {\bibfnamefont {K.}~\bibnamefont {Wakabayashi}}, \
  and\ \bibinfo {author} {\bibfnamefont {C.-H.}\ \bibnamefont {Lam}},\
  }\href@noop {} {\bibfield  {journal} {\bibinfo  {journal} {ArXiv}\ ,\
  \bibinfo {pages} {1511.07776}} (\bibinfo {year} {2015})}\BibitemShut
  {NoStop}%
\bibitem [{\citenamefont {Barnes}(2006)}]{1464-4258-8-4-S06}%
  \BibitemOpen
  \bibfield  {author} {\bibinfo {author} {\bibfnamefont {W.~L.}\ \bibnamefont
  {Barnes}},\ }\href {http://stacks.iop.org/1464-4258/8/i=4/a=S06} {\bibfield
  {journal} {\bibinfo  {journal} {J. Optics A: Pure and Appl. Optics}\ }\textbf
  {\bibinfo {volume} {8}},\ \bibinfo {pages} {S87} (\bibinfo {year}
  {2006})}\BibitemShut {NoStop}%
\bibitem [{Note1()}]{Note1}%
  \BibitemOpen
  \bibinfo {note} {Supplemental Online Information.}\BibitemShut {Stop}%
\bibitem [{\citenamefont {West}\ \emph {et~al.}(2010)\citenamefont {West},
  \citenamefont {Ishii}, \citenamefont {Naik}, \citenamefont {Emani},
  \citenamefont {Shalaev},\ and\ \citenamefont {Boltasseva}}]{west}%
  \BibitemOpen
  \bibfield  {author} {\bibinfo {author} {\bibfnamefont {P.}~\bibnamefont
  {West}}, \bibinfo {author} {\bibfnamefont {S.}~\bibnamefont {Ishii}},
  \bibinfo {author} {\bibfnamefont {G.}~\bibnamefont {Naik}}, \bibinfo {author}
  {\bibfnamefont {N.}~\bibnamefont {Emani}}, \bibinfo {author} {\bibfnamefont
  {V.}~\bibnamefont {Shalaev}}, \ and\ \bibinfo {author} {\bibfnamefont
  {A.}~\bibnamefont {Boltasseva}},\ }\href@noop {} {\bibfield  {journal}
  {\bibinfo  {journal} {Laser \& Photonics Rev.}\ }\textbf {\bibinfo {volume}
  {4}},\ \bibinfo {pages} {795} (\bibinfo {year} {2010})}\BibitemShut {NoStop}%
\bibitem [{\citenamefont {Johnson}\ and\ \citenamefont
  {Christy}(1972)}]{johnson1972}%
  \BibitemOpen
  \bibfield  {author} {\bibinfo {author} {\bibfnamefont {P.~B.}\ \bibnamefont
  {Johnson}}\ and\ \bibinfo {author} {\bibfnamefont {R.~W.}\ \bibnamefont
  {Christy}},\ }\href@noop {} {\bibfield  {journal} {\bibinfo  {journal} {Phys.
  Rev. B}\ }\textbf {\bibinfo {volume} {6}},\ \bibinfo {pages} {4370} (\bibinfo
  {year} {1972})}\BibitemShut {NoStop}%
\bibitem [{\citenamefont {Bliokh}\ \emph {et~al.}(2015)\citenamefont {Bliokh},
  \citenamefont {Smirnova},\ and\ \citenamefont {Nori}}]{Bliokh26062015}%
  \BibitemOpen
  \bibfield  {author} {\bibinfo {author} {\bibfnamefont {K.~Y.}\ \bibnamefont
  {Bliokh}}, \bibinfo {author} {\bibfnamefont {D.}~\bibnamefont {Smirnova}}, \
  and\ \bibinfo {author} {\bibfnamefont {F.}~\bibnamefont {Nori}},\ }\href@noop
  {} {\bibfield  {journal} {\bibinfo  {journal} {Science}\ }\textbf {\bibinfo
  {volume} {348}},\ \bibinfo {pages} {1448} (\bibinfo {year}
  {2015})}\BibitemShut {NoStop}%
\bibitem [{\citenamefont {Drachev}\ \emph {et~al.}(2008)\citenamefont
  {Drachev}, \citenamefont {Chettiar}, \citenamefont {Kildishev}, \citenamefont
  {Yuan}, \citenamefont {Cai},\ and\ \citenamefont {Shalaev}}]{drachev2008}%
  \BibitemOpen
  \bibfield  {author} {\bibinfo {author} {\bibfnamefont {V.~P.}\ \bibnamefont
  {Drachev}}, \bibinfo {author} {\bibfnamefont {U.~K.}\ \bibnamefont
  {Chettiar}}, \bibinfo {author} {\bibfnamefont {A.~V.}\ \bibnamefont
  {Kildishev}}, \bibinfo {author} {\bibfnamefont {H.-K.}\ \bibnamefont {Yuan}},
  \bibinfo {author} {\bibfnamefont {W.}~\bibnamefont {Cai}}, \ and\ \bibinfo
  {author} {\bibfnamefont {V.~M.}\ \bibnamefont {Shalaev}},\ }\href@noop {}
  {\bibfield  {journal} {\bibinfo  {journal} {Optics express}\ }\textbf
  {\bibinfo {volume} {16}},\ \bibinfo {pages} {1186} (\bibinfo {year}
  {2008})}\BibitemShut {NoStop}%
\bibitem [{\citenamefont {Rhodes}\ \emph {et~al.}(2006)\citenamefont {Rhodes},
  \citenamefont {Franzen}, \citenamefont {Maria}, \citenamefont {Losego},
  \citenamefont {Leonard}, \citenamefont {Laughlin}, \citenamefont {Duscher},\
  and\ \citenamefont {Weibel}}]{rhodes2006}%
  \BibitemOpen
  \bibfield  {author} {\bibinfo {author} {\bibfnamefont {C.}~\bibnamefont
  {Rhodes}}, \bibinfo {author} {\bibfnamefont {S.}~\bibnamefont {Franzen}},
  \bibinfo {author} {\bibfnamefont {J.-P.}\ \bibnamefont {Maria}}, \bibinfo
  {author} {\bibfnamefont {M.}~\bibnamefont {Losego}}, \bibinfo {author}
  {\bibfnamefont {D.~N.}\ \bibnamefont {Leonard}}, \bibinfo {author}
  {\bibfnamefont {B.}~\bibnamefont {Laughlin}}, \bibinfo {author}
  {\bibfnamefont {G.}~\bibnamefont {Duscher}}, \ and\ \bibinfo {author}
  {\bibfnamefont {S.}~\bibnamefont {Weibel}},\ }\href@noop {} {\bibfield
  {journal} {\bibinfo  {journal} {Journal of Applied Physics}\ }\textbf
  {\bibinfo {volume} {100}},\ \bibinfo {pages} {054905} (\bibinfo {year}
  {2006})}\BibitemShut {NoStop}%
\bibitem [{\citenamefont {Losego}\ \emph {et~al.}(2009)\citenamefont {Losego},
  \citenamefont {Efremenko}, \citenamefont {Rhodes}, \citenamefont {Cerruti},
  \citenamefont {Franzen},\ and\ \citenamefont {Maria}}]{losego2009}%
  \BibitemOpen
  \bibfield  {author} {\bibinfo {author} {\bibfnamefont {M.~D.}\ \bibnamefont
  {Losego}}, \bibinfo {author} {\bibfnamefont {A.~Y.}\ \bibnamefont
  {Efremenko}}, \bibinfo {author} {\bibfnamefont {C.~L.}\ \bibnamefont
  {Rhodes}}, \bibinfo {author} {\bibfnamefont {M.~G.}\ \bibnamefont {Cerruti}},
  \bibinfo {author} {\bibfnamefont {S.}~\bibnamefont {Franzen}}, \ and\
  \bibinfo {author} {\bibfnamefont {J.-P.}\ \bibnamefont {Maria}},\ }\href@noop
  {} {\bibfield  {journal} {\bibinfo  {journal} {Journal of Applied Physics}\
  }\textbf {\bibinfo {volume} {106}},\ \bibinfo {pages} {024903} (\bibinfo
  {year} {2009})}\BibitemShut {NoStop}%
\bibitem [{\citenamefont {Prade}\ \emph {et~al.}(1991)\citenamefont {Prade},
  \citenamefont {Vinet},\ and\ \citenamefont {Mysyrowicz}}]{prada}%
  \BibitemOpen
  \bibfield  {author} {\bibinfo {author} {\bibfnamefont {B.}~\bibnamefont
  {Prade}}, \bibinfo {author} {\bibfnamefont {J.~Y.}\ \bibnamefont {Vinet}}, \
  and\ \bibinfo {author} {\bibfnamefont {A.}~\bibnamefont {Mysyrowicz}},\
  }\href@noop {} {\bibfield  {journal} {\bibinfo  {journal} {Phys. Rev. B}\
  }\textbf {\bibinfo {volume} {44}},\ \bibinfo {pages} {13556} (\bibinfo {year}
  {1991})}\BibitemShut {NoStop}%
\end{thebibliography}%

\end{document}